\documentclass[pra,aps,twocolumn,showpacs,epsfig]{revtex4}
\usepackage{color}
\usepackage{graphicx,amssymb,epsfig,epsf,amsmath}
\begin{document}
\title{
Condensate fraction and critical temperature of interacting Bose gas in anharmonic trap.
}
\author{
Sudip Kumar Haldar$^{1}$,
\footnote{e-mail: sudip\_cu@rediffmail.com}, 
Barnali Chakrabarti$^{2}$, Satadal Bhattacharya$^{3}$, Tapan Kumar Das$^4$
}

\affiliation{
$^{1}$Department of Physics, Lady Brabourne College, P-$\frac{1}{2}$ Suhrawardi Avenue, Kolkata-700017, India.\\
$^{2}$Department of Physics, Presidency University, 86/1, College Street, Kolkata-700073, India. \\
$^{3}$Department of Physics, Scottish Church College, 1\& 3, Urquhart square, Kolkata-700006, India.\\
$^{4}$Department of Physics, Calcutta University, 92, A. P. C. Road, Kolkata-700009, India.}

\begin{abstract}
By using a correlated many body method and using the realistic van der Waals potential we study several 
statistical measures like the specific heat, transition temperature and the condensate fraction of the 
interacting Bose gas trapped in an anharmonic potential. As the quadratic plus a quartic confinement makes the trap
more tight, the transition temperature increases which makes more favourable condition to achieve Bose-Einstein 
condensation (BEC) experimentally. BEC in 3D isotropic harmonic potential is also critically studied, the correction to the
critical temperature due to finite number of atoms and also the correction due to inter-atomic interaction are calculated
by the correlated many-body method. Comparison and discussion with the mean-field  results are presented. 
\end{abstract}
\maketitle
\section{Introduction}
The observations of Bose-Einstein condensation (BEC) in trapped atomic gases have stimulated a new interest in the experimental 
and theoretical research of quantum gases~\cite{Dalfovo}. BEC is characterized by a macroscopic occupation in the ground state below 
some critical temperature $T_c$~\cite{Huang, Stringari}. 

The experimental situation is quite different from the ideal Bose gas (IBG) which is generally treated within the grand canonical 
ensemble~\cite{Huang}. The confining trap also greatly influences the condensate properties~\cite{Dalfovo, Stringari}. 
Although the atomic cloud is extremely dilute, the 
interatomic forces strongly influence the energy of the condensate cloud. The elementary excitations are also seriously affected 
by the interaction. The question of how the two-body forces affect the thermodynamic properties is also a subject of great interest. 
By using mean-field approach, based on the Popov approximation the condensate fraction and the critical temperature of the trapped interacting bosons has been calculated~\cite{Giorgini}. A 
significant decrease of the condensate fraction and the same of the critical temperature for repulsive BEC with respect to the 
prediction of non-interacting model was observed. Path-integral Monte carlo method has also been employed for 1D Bose gas~\cite{Pearson1}. The 
possibility of observation of low-dimensional BEC of hard sphere bosons~\cite{Pearson} and the calculation of thermodynamic properties have also been discussed in several theoretical investigations~\cite{Giorgini, Pilati, Borrmann, MK, Shi}.

The aim of our present study is to calculate the critical temperature and condensate fraction for BEC in anharmonic trap. The trap 
potential is modelled as $V(r) = \frac{1}{2}m \omega^2 r^2 + \lambda r^4$ where $\lambda$ is a controllable parameter.  For 
$\lambda>0$ the quartic confinement becomes more tight. Since the transition temperature $T_c$ increases in the tight trap (with 
increase in $\lambda$), BEC may be achieved more favorably in the tight trap. In the experiments this is achieved by tuning the 
laser intensity. It has been demonstrated that starting with a dilute interacting gas at a 
temperature higher than the transition temperature phase-space density required for BEC can be achieved 
by varying the shape of the trap adiabatically~\cite{Ketterle, Pinkse, DStamper}. 
This allows to study the kinetics of condensate formation and 
effect of interaction and finite number of particles on the phase transition. 
In the experiments, local phase-space density of an ultracold gas was adiabatically changed to obtain BEC in a reversible manner by using a combination of magnetic and optical forces~\cite{DStamper}.  
Adiabatic cooling has also been studied in power-law potential~\cite{Bagnato} but such processes in a harmonic plus quartic potential is yet to be studied.
In the present work we start with $N$ particle Schr\"odinger equation which represent the $N$ interacting bosons in the external 
trap. The interatomic interaction is chosen as van der Waals potential with a hard core of very short range and long-range 
attractive tail. We expand the condensate wave function in the basis of two-body Faddeev component. 
The correlated basis function keeps all possible two-body correlations which facilitates us to study the beyond mean-field effect.
It is to be noted that the number of bosons in the condensate varies from few hundred to several thousands which is far below the 
thermodynamic limit.

The paper is organised as follows. In sec. II we briefly describe the methodology. We discuss our results in sec. III and draw our
conclusions in sec. IV. 
\section{Methodology} 
\subsection{Potential harmonic expansion (CPHEM) method}

In our present study we adopt an ab-initio many-body technique which is known as potential harmonic expansion method (PHEM). PHEM uses 
a truncated two-body basis set which keeps all possible two-body 
correlation~\cite{Fabre} and hence it is an improvement over the mean-field approximation. The 
potential harmonic expansion method with an additional short range 
correlation function, called CPHEM, has already been established as 
a very successful and useful technique for the study of dilute BEC~\cite{Anindya1,Sudip1,Anindya2,Sudip2,Sudip3}. Here 
we briefly describe the technique. 
Details can be found in our earlier work~\cite{Tapan, Das, Kundu}.

The Hamiltonian for a system of $N$ identical bosons interacting via two-body potential $V(\vec{r}_{ij}) = V(\vec{r}_{i}-\vec{r}_{j})$ and confined in an external trap has the form
\begin{equation}
H=-\frac{\hbar^2}{2m}\sum_{i=1}^{N} \nabla_{i}^{2} 
+ \sum_{i=1}^{N} V_{trap}(\vec{r}_{i}) 
+\displaystyle{\sum_{i,j>i}^{N}} V(\vec{r}_{i}-\vec{r}_{j})\cdot
\end{equation}
Here $m$ is the mass of each bosons. After eliminating the center of mass motion by using standard Jacobi coordinates~\cite{Fabre,Ballot,MFabre}, defined by
\begin{equation}
 \vec{\zeta}_{i}=\sqrt{\frac{2i}{i+1}}(\vec{r}_{i+1}-
\frac{1}{i}\sum_{j=1}^{i} \vec{r}_j) \hspace*{.5cm}
 (i=1,...,\mathcal{N}); (\mathcal{N}=N-1),
\end{equation}
we obtain the Hamiltonian of the relative motion as
\begin{equation}
H=-\frac{\hbar^{2}}{m}\sum_{i=1}^{\mathcal{N}} 
\nabla_{\zeta_{i}}^{2}+V_{trap} + V_{int}
(\vec{\zeta}_{1}, ..., \vec{\zeta}_{\mathcal{N}})\hspace*{.1cm}, 
\end{equation}
Here $V_{int}$ is the sum of all pair-wise interactions expressed in terms of the Jacobi vectors. The Hyperspherical harmonic expansion method (HHEM) is a convenient  {\it ab-initio} many-body tool which includes all correlations~\cite{Ballot}. However 
it can not be applied to a typical BEC containing few thousands to few millions of atoms due to large degeneracy of the HH basis. To simplify things now we explore the features of experimentally achieved BEC. In a typical laboratory BEC the interparticle separation is very large compared to the range of interatomic interaction and hence the probabilities of three and higher body collision is negligible. So we can safely ignore the effect of three-body and higher-body correlation and can keep only the two-body  correlation.
It permits us to decompose the total wave function $\Psi$ into two-body Faddeev component 
for the interacting $(ij)$ pair as 
\begin{equation}
\Psi=\sum_{i,j>i}^{N}\phi_{ij}(\vec{r}_{ij},r)\hspace*{.1cm}\cdot
\end{equation}
It is worth to note that $\phi_{ij}$ is a function of two-body 
separation ($\vec{r}_{ij}$) only and also includes the global 
hyperradius $r$, which is defined as
$r = \sqrt{\sum_{i=1}^{\mathcal{N}}\zeta_{i}^{2}}$.
Thus the effect of two-body correlation comes 
through the two-body interaction in the expansion basis. $\phi_{ij}$ 
is symmetric under the exchange operator $P_{ij}$ for bosonic atoms and satisfy the 
Faddeev equation
\begin{equation}
\left[T+V_{trap}-E_R\right]\phi_{ij}
=-V(\vec{r}_{ij})\sum_{kl>k}^{N}\phi_{kl}
\end{equation}
where $T$ is the total kinetic energy operator. Operating $\sum_{i,j>i}$ on both 
sides of equation (4), we get back the original Schr\"odinger equation. 
In this approach, we assume that when ($ij$) pair interacts, the rest are just inert spectators. 
Thus the total hyperangular momentum 
quantum number as also the orbital angular momentum of the whole system 
is contributed by the interacting pair only. Next  the  $(ij)$th Faddeev 
component is expanded in the set of potential harmonics (PH) (which is 
a subset of hyperspherical harmonic (HH) basis and sufficient for the expansion of $V(\vec {r}_{ij})$) 
appropriate for the ($ij$) partition as 
\begin{equation}
\phi_{ij}(\vec{r}_{ij},r)
=r^{-(\frac{3{\mathcal{N}}-1}{2})}\sum_{K}{\mathcal P}_{2K+l}^{lm}
(\Omega_{\mathcal{N}}^{ij})u_{K}^{l}(r) \hspace*{.1cm}\cdot
\end{equation}
$\Omega_{\mathcal{N}}^{ij}$ denotes the full set of hyperangles in the $3\mathcal{N}$-dimensional 
space corresponding to the $(ij)$th  interacting pair and 
${\mathcal P}_{2K+l}^{lm}(\Omega_{\mathcal{N}}^{ij})$ is called the PH basis. It 
has an analytic expression:
\begin{equation}
{\mathcal P}_{2K+l}^{l,m} (\Omega_{\mathcal{N}}^{(ij)}) =
Y_{lm}(\omega_{ij})\hspace*{.1cm} 
^{(\mathcal{N})}P_{2K+l}^{l,0}(\phi) {\mathcal Y}_{0}(D-3) ;\hspace*{.5cm}D=3{\mathcal{N}} ,
\end{equation}
${\mathcal Y}_{0}(D-3)$ is the HH of order zero in 
the $(3{\mathcal{N}}-3)$ dimensional space spanned by $\{\vec{\zeta}_{1}, ...,
\vec{\zeta}_{\mathcal{N}-1}\}$ Jacobi vectors; $\phi$ is the hyperangle given by
$r_{ij}$ = $r\hspace*{0.1cm} cos\phi$. For the remaining $(\mathcal{N}-1)$
noninteracting bosons we define hyperradius as
\begin{eqnarray}
 \rho_{ij}& = &\sqrt{\sum_{K=1}^{\mathcal{N}-1}\zeta_{K}^{2}}\nonumber\\
          &= &r \sin\phi \hspace*{.01 cm}\cdot
\end{eqnarray}
such that $r^2=r_{ij}^2+\rho_{ij}^2$ and $r$ represents the global 
hyperradius of the condensate. The set of $(3\mathcal{N}-1)$ quantum 
numbers of HH is now reduced to {\it only} $3$ as for the $(\mathcal{N}-1)$ 
non-interacting pair we have
\begin{eqnarray}
l_{1} = l_{2} = ...=l_{{\mathcal{N}}-1}=0,   & \\
m_{1} = m_{2}=...=m_{{\mathcal{N}}-1}=0,  &   \\
n_{2} = n_{3}=...n_{{\mathcal{N}}-1} = 0, & 
\end{eqnarray}
whereas for the interacting pair $l_{\mathcal{N}} = l$, $m_{\mathcal{N}} = m$ and  $n_{\mathcal{N}} = K$.
Thus the $3\mathcal{N}$ dimensional Schr\"odinger equation is effectively reduced
to a four dimensional equation with the relevant set of quantum 
numbers: hyperradius $r$, orbital angular momentum quantum number $l$,
azimuthal quantum number $m$ and grand orbital quantum number $2K+l$
 for any $\mathcal{N}$.
Substituting in Eq(4) and projecting on a particular PH, a set of 
coupled differential equation (CDE) for the partial wave $u_{K}^{l}(r)$
is obtained
\begin{equation}
\begin{array}{cl}
&\Big[-\frac{\hbar^{2}}{m} \frac{d^{2}}{dr^{2}} +
V_{trap}(r) + \frac{\hbar^{2}}{mr^{2}}
\{ {\cal L}({\cal L}+1) \\
&+ 4K(K+\alpha+\beta+1)\}-E_R\Big]U_{Kl}(r)\\
+&\displaystyle{\sum_{K^{\prime}}}f_{Kl}V_{KK^{\prime}}(r)
f_{K^{\prime}l}
U_{K^{\prime}l}(r) = 0
\hspace*{.1cm},
\end{array}
\end{equation}\\
where ${\mathcal L}=l+\frac{3N-6}{2}$, $U_{Kl}=f_{Kl}u_{K}^{l}(r)$, 
$\alpha=\frac{3N-8}{2}$ and $\beta=l+1/2$.\\
$f_{Kl}$ is a constant and represents the overlap of the PH for
interacting partition with the sum of PHs corresponding  to all 
partitions~\cite{MFabre}.
The potential matrix element $V_{KK^{\prime}}(r)$ is given by
\begin{equation}
V_{KK^{\prime}}(r) =  
\int P_{2K+l}^{lm^*}(\Omega_{\mathcal{N}}^{ij}) 
V\left(r_{ij}\right)
P_{2K^{\prime}+1}^{lm}(\Omega_{\mathcal{N}}^{ij}) d\Omega_{\mathcal{N}}^{ij} 
\hspace*{.1cm}\cdot
\end{equation}

\subsection{Introduction of additional short range correlation}

In the mean-field GP theory the inter-atomic interaction is modelled by the $\delta$-potential whose strength is
characterized by the $s$-wave scattering length $a_{s}$ only. 
Thus it completely ignores the detailed structure of the interaction potential. However we have already reported that shape-dependent interaction potential is indeed necessary for the description of experimental BEC~\cite{PRA2008}. So in our present study, we model the inter-atomic interaction by a realistic potential such as 
van der Waals potential with a hard core at short separation and having an attractive $-\frac{1}{r_{ij}^6}$ tail at 
large separation. Due to the use of such realistic the two-body potential with
detailed structure, it becomes necessary to incorporate
an additional short range correlation function in the PH basis. This short 
range behavior is well represented by a hard core of radius $r_c$ and we 
calculate the two-body wave function $\eta(r_{ij})$ by solving the 
zero-energy two-body Schrodinger equation
\begin{equation}
-\frac{\hbar^2}{m}\frac{1}{r_{ij}^2}\frac{d}{dr_{ij}}\left(r_{ij}^2
\frac{d\eta(r_{ij})}{dr_{ij}}\right)+V(r_{ij})\eta(r_{ij})=0
\hspace*{.1cm}\cdot
\end{equation} 
This zero-energy two-body wave function $\eta(r_{ij})$  
represents the short range behavior of $\phi_{ij}$ quite well because in the actual
experimental BEC, the energy of the interacting pair is negligible 
compared with the depth of the interatomic potential. Thus it is 
taken as the two-body correlation function in the PH expansion 
basis. The value of $r_c$ is adjusted to obtain the expected value of $a_{s}$~\cite{Anasua}. We 
introduce this as a short-range correlation function in the expansion 
basis. This also greatly improves the convergence rate of the 
PH basis and we call it as correlated Potential Harmonic expansion 
method (CPHEM). This replaces Eq(5) by 
\begin{equation}
\phi_{ij}(\vec{r}_{ij},r)
=r^{-(\frac{3\mathcal{N}-1}{2})}\sum_{K}{\mathcal P}_{2K+l}^{lm}
(\Omega_{\mathcal{N}}^{ij})u_{K}^{l}(r) \eta(r_{ij}) \hspace*{.1cm}\cdot
\end{equation}
and the correlated PH (CPH) basis is given by
\begin{equation}
[{\mathcal P}_{2K+l}^{l,m} (\Omega_{\mathcal{N}}^{(ij)})]_{correlated} =
 {\mathcal P}_{2K+l}^{l,m} (\Omega_{\mathcal{N}}^{(ij)}) \eta(r_{ij}) ,
\end{equation}
The correlated potential matrix $V_{KK^{\prime}}(r)$ is now given by
\begin{equation}
\begin{array}{cl}
&V_{KK^{\prime}}(r) =(h_{K}^{\alpha\beta} h_{K^{\prime}}^
{\alpha\beta})^{-\frac{1}{2}}\times \\
&\int_{-1}^{+1} \{P_{K}^{\alpha\beta}(z) 
V\left(r\sqrt{\frac{1+z}{2}}\right)
P_{K^{\prime}}^{\alpha \beta}(z)\eta\left(r\sqrt{\frac{1+z}{2}}\right)
W_{l}(z)\} dz \hspace*{.1cm}\cdot
\end{array}
\end{equation}
Here $P_{K}^{\alpha\beta}(z)$ is the Jacobi polynomial, and its 
norm and weight function are $h_{K}^{\alpha\beta}$ and $W_{l}(z)$   
respectively~\cite{Abramowitz}.

This is to be noted that the inclusion of $\eta(r_{ij})$ makes the PH basis 
non-orthogonal. One may surely use the standard procedure for handling 
non-orthogonal basis. However in the present calculation we have 
checked that $\eta(r_{ij})$ differs from a constant value only by small 
amount and the overlap $\Big< {\mathcal P}_{2K+l}^{l,m} (\Omega_{\mathcal{N}}^{(ij)})|{\mathcal P}_{2K+l}^{l,m} (\Omega_{\mathcal{N}}^{(kl)})\eta(r_{kl})\Big>$ is quite small. Thus we can ignore its derivatives and obtain the Eq(12) 
approximately when the correlated potential matrix is calculated by Eq(17). This implies that the effective two-body interaction now becomes $V(r_{ij})\eta(r_{ij})$. Physically since the interacting atoms of the condensate have very low energy, they do not come close enough to see the actual interatomic interaction. In the zero-energy limit, the scattering cross-section becomes $4\pi a_s^2$ and the effective interaction seen by the atoms is governed by $a_s$ through $\eta(r_{ij})$.

\section{Numerical results}
\subsection{Choice of two-body potential and solution of 
many-body effective potential}

The interatomic potential has been chosen as the van der Waals 
potential with a hard core of radius $r_{c}$, viz., 
$V(r_{ij}) = \infty$ for $r_{ij} \le r_{c}$ and $= -\frac{C_6}{r_{ij}^6}$
for $r_{ij} > r_{c}$. $C_6$ is known for a 
specific atom and for $^{87}$Rb atom $C_6= 6.4898 \times 10^{-11}$ o.u.~\cite{Pethick}. In the limit of $C_6 \rightarrow 0$, the potential becomes a hard sphere and the cutoff radius exactly coincides with the $s$-wave scattering length $a_{s}$. In our choice of two-body potential we tune $r_c$ to reproduce the experimental scattering length. As we decrease $r_c$, $a_{sc}$ decreases and at a particular critical value of $r_c$  it passes through $- \infty$ to $\infty$~\cite{Kundu}. For repulsive BEC, $r_c$ is so chosen that it corresponds to the one-node in the two-body wave function. For our present calculation we consider $^{87}$Rb atoms with $a_{s}$ = $0.00433$ o.u. which mimics the JILA experiments~\cite{Anderson}. We tune $r_c$ to obtain the desired value of $a_s$ and our chosen value of $r_c$ is $1.121 \times 10^{-3}$ o.u.  With these set of parameters we solve the coupled differential equation by hyperspherical adiabatic approximation (HAA)~\cite{Coelho}. In HAA, we assume that the hyperradial motion is slow compared to the 
hyperangular motion and the potential matrix together with the 
hypercentrifugal repulsion is diagonalized for a fixed value of $r$. 
Thus the effective potential for the hyperradial motion is obtained 
as a parametric function of $r$. We choose the lowest eigen potential 
$\omega_0(r)$ as the effective potential in which the condensate moves
collectively. The energy and wave function of the condensate are 
finally obtained by solving the adiabatically separated hyperradial equation 
in the extreme adiabatic approximation (EAA)
\begin{equation}
 \left[-\frac{\hbar^{2}}{m}\frac{d^{2}}{dr^{2}}+\omega_{0}(r)-E_{R}
\right]\zeta_{0}(r)=0\hspace*{.1cm},
\label{EAA}
\end{equation}
subject to approximate boundary condition on $\zeta_0(r)$.
For our numerical calculation we fix $l=0$ and truncate the CPH basis 
to a maximum value $K=K_{max}$ requiring proper convergence. 

 To study the thermodynamics of the condensate we need to calculate a large number of energy levels in the effective potential. The ground state in the effective eigen potential $\omega_0(r)$ corresponds to the ground state energy $E_{00}$ of the condensate. Hyperradial excitations for $l$=0 in this effective eigen potential $\omega_0(r)$ corresponds to the breathing mode. Similarly the surface mode excitations with $l>0$ can be calculated as hyperradial excitations in the eigen potential $\omega_{l}(r)$ corresponding to different values of $l$. However, for $l$ $\neq$ 0, a large inaccuracy is involved in the calculation of off-diagonal potential matrix and numerical computation becomes very slow. However, the main contribution to the potential matrix comes from the diagonal hypercentrifugal term and we disregard the off-diagonal matrix element for $l>0$. Thus we get the effective potential $\omega_{l}(r)$ in the hyperradial space for $l\neq 0$ by adding the hypercentrifugal term corresponding to a particular value of $l$ with the potential matrix element for $l=0$. The ground state in this potential is the ground state of the $l$th surface mode $E_{0l}$ and $n$th radial excitations provide the $E_{nl}$. Finally we add the center of mass energy $1.5 \hbar \omega$ to each energy eigen value to obtain the actual energy states.  \\
 
Before using the energy spectrum for the calculation of thermodynamic properties, it needs at least some qualitative and quantitative discussion on how good our approximation for the calculation of higher order excitations is. There are many approximation methods which calculate the low-lying collective excitations and also the higher multipolarities~\cite{Dalfovo}. All these basically use the uncorrelated mean-feld theory and the hydrodynamic (HD) model. It is well known that the excited states at high energy are expected to have single particle nature whereas the low-lying excitations are of collective nature. However the transition from 
collective to single-particle excitations for trapped interacting bosons may be systematically affected by the interatomic correlation and finite size effect. The HD model is good for a large number of bosons in the trap in the Thomas-Fermi limit and in the spherical trap the eigenfrequencies are calculated using the analytic formula~\cite{Dalfovo, Dalfovo1}
\begin{equation}
\omega(n,l)=\omega_{ho}(2n^2+2nl+3n+l)^{1/2} .
\label{eq.HD}
\end{equation} 
Here $n$ is the radial quantum number and $l$ is the angular momentum quantum number. $n=0$ corresponds to the surface mode
excitation, whereas monopole oscillation $\omega_M$ corresponds to $n=1$ and $l=0$.
Note that Eq.~(\ref{eq.HD}) has the dependence on the radial nodes and angular momentum but is independent of particle number,
whereas in the extreme case of noninteracting harmonic oscillator (HO) model
\begin{equation}
\omega(n,l)= \omega_{ho}(2n+l)
\label{eq.HO}
\end{equation}

\begin{figure}[hbpt]
\vspace{-10pt}
\centerline{
\hspace{-3.3mm}
\rotatebox{0}{\epsfxsize=8.8cm\epsfbox{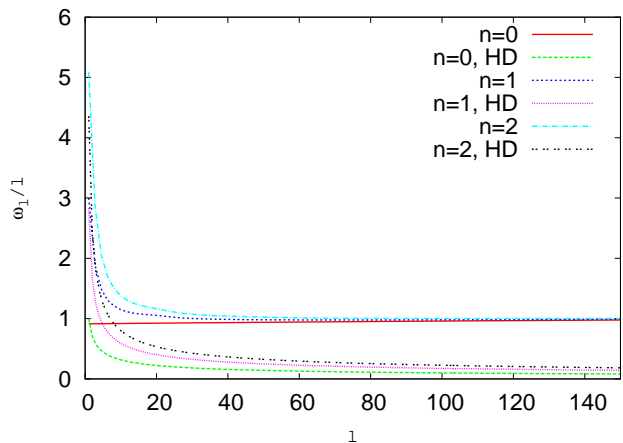}}}
\caption{(color online) Plot of the different excitation mode ($n$=0,1,2) frequencies $\omega_l/l$ of the BEC for $N=200$.}
\label{fig.excitation-200}
\end{figure} 

\begin{figure}[hbpt]
\vspace{-10pt}
\centerline{
\hspace{-3.3mm}
\rotatebox{0}{\epsfxsize=8.8cm\epsfbox{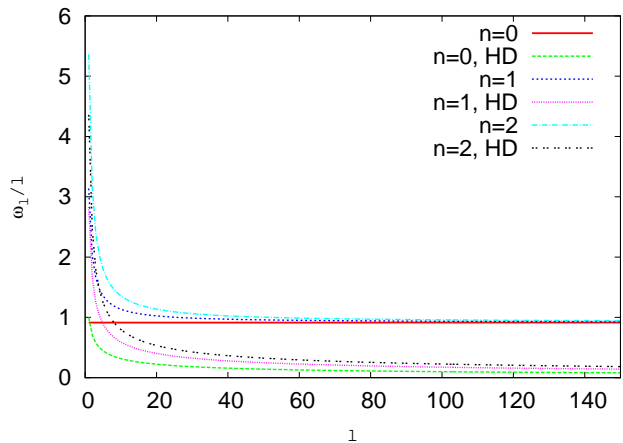}}}
\caption{(color online) Plot of the different excitation mode ($n$=0,1,2) frequencies $\omega_l/l$ of the BEC for $N=11000$.}
\label{fig.excitation-11k}
\end{figure}

Thus in general the HD model is expected to be accurate for the low energy excitations of the system and ideal gas is expected
 to be valid for high excitation energies. Thus our many-body calculation basically connects the two extreme limits. However it 
 additionally incorporate the effect of finite-size  and inter-atomic correlation. We calculate the frequency of the surface modes and compare with the predictions of HD model. In Fig~\ref{fig.excitation-200} we show the evolution of several excitation frequencies ($\omega_{l}/l$ in unit of $\omega_{ho}$) for the different modes ($n=0,1,2$) as a function of their angular momentum $l$ for $N=200$ atoms. It is mentioned already that the effect of interaction is particularly important for the surface mode $(n=0)$. In the same figure the hydrodynamic predictions is also shown. It shows that our many-body results first follow the HD curve and then very soon it starts to deviate. However deviation is larger for $n=0$ and it generally starts to decrease with larger $n$. Another important point is the asymptotic limit when the HD prediction assymptotically goes to zero for large $l$. The many-body results approach asymptotically the correct non-interacting limit $\frac{\omega_l}{l} = \omega_{ho}=1$ o.u.. A similar observation is also made in Ref.~\cite{Dalfovo1}, where the excitation spectrum of $N=10000$ $^{87}$Rb atoms in a spherical trap was calculated using Hartree-Bogoliubov theory and compared with the HD prediction. For larger particles ($N=11000$), we observe (Fig.~\ref{fig.excitation-11k}) the same qualitative nature, whereas $\frac{\omega_l}{l}$ for different surface modes become closer to the HD prediction as expected. 
 
Also from Fig.~\ref{fig.excitation-200} and Fig.~\ref{fig.excitation-11k} we observe that for larger $l$ the frequencies ($\omega_{l}/l$) of different modes of excitations asymptotically approach the harmonic oscillator frequency $\omega_{ho}$. 
It indicates that 
for the higher excitations, dispersion relation becomes independent of the two-body interaction. In our many-body theory this can be justified as follows. For the higher excited states, the average size of the condensate increases and the atoms tend to be pushed away from each other. This is also consistent with the findings from the Hartree-Bogoliubov theory~\cite{Dalfovo} where the Bogoliubov-type spectrum exhibits states of single-particle nature localized near the surface of the condensate. Surface mode excitations in the Bogoliubov-type spectrum can not be of collective nature when its wave length is smaller than the surface thickness $d$. This condition is satisfied when $l$ exceeds a critical value
$\sim N^{4/15}$~\cite{Dalfovo} and the HO model becomes a better approximation for individual excited states with larger $l$ values. However low-lying states are strongly collective, whose many-body effects are properly taken care of by the CPHEM. For calculation of thermodynamic quantities, sums over many excited states, in addition to the low-lying ones, are needed. This masks the effect of single particle nature, as one can see in a typical plot of number of states $N(\epsilon)$ with energy $\leq \epsilon$ against $\epsilon$ in Fig.~21 of Ref.~\cite{Dalfovo}.
    
\subsection{Bose-Einstein condensation in three dimensional isotropic harmonic trap}

In this section we will discuss Bose-Einstein condensation in three dimensional harmonic trap and move on to the anharmonic trap in the next section. The system of $N$ non-interacting bosons is analytically treated in the grand-canonical ensemble and in the thermodynamic limit there will be a phase transition at a critical temperature $T_c$. In three dimension the critical temperature 
$T_c$ is finite and the phase-transition is of first order. However the BEC experiments are also carried out with a finite number of 
atoms, thus the thermodynamic argument does not apply here. At $T=T_c$, the chemical potential $\mu =0$ whereas for finite number of 
particles, $\mu$ and other thermodynamic quantities are smooth function 
of temperature. Thus the possibility of phase transition is ruled out. Therefore one can not define a critical temperature for a 
system of finite number of bosons. Although there is no phase transition for finite size systems, there is still a macroscopic 
number of particles in the ground state at low enough temperature. Thus
it is convenient to define a transition temperature in the same way as the critical temperature in the 
thermodynamic limit.  Also at low temperature and in presence of confining potential the interatomic 
force becomes important even for very dilute atomic cloud. Thus the transition temperature deviates 
from the critical temperature of ideal gas. The effect of interaction on the thermodynamic properties 
of the gas has also been pointed out which signals the breakdown of the Bogoliubov theory~\cite{Bogoliubov} 
in the calculation of collective excitation. Thus the weakly interacting finite size condensate is of considerable interest.

By using the mean-field approach, based on the Popov approximations~\cite{Dalfovo, Pethick}, 
the temperature dependence of the condensate 
with repulsive forces is calculated.  It is shown that the first correction to the critical temperature due to finite number
of atoms obey~\cite{Grossmann} 
\begin{equation}
\frac{\delta T_c}{T_c^0}{\Big |}_{finite-size} = -0.73 \frac{\bar{\omega}}{\omega} N^{-1/3}
\label{eq.finite-correction}
\end{equation}    
where $\bar{\omega}$ is the mean frequency of a non-isotropic harmonic trap. Mean-field theory also predicts a relative shift in critical temperature due to inter-atomic interaction as~\cite{Giorgini} 
\begin{equation}
\frac{\delta T_c}{T_c^0}{\Big |}_{int} = -1.33a_sN^{1/6}
\label{eq.int-shift}
\end{equation} 

In this paper we utilize the correlated potential harmonic expansion technique as discussed in Sec. II. The correlated 
basis function offers to study the beyond mean-field effect and to calculate various thermodynamic properties more accurately. 
As we assume that the inter-atomic correlations beyond two-body can safely be ignored in our picture, it reduces computational 
difficulty. We can handle quite few to very large number of atoms by a single technique. Although very recently 
we have employed this technique to calculate several thermodynamic properties of weakly interacting Bose gas, the 
finite-size effect and deviation due to inter-atomic interaction was not studied~\cite{Sanchari1, Sanchari2, Satadal}. 
Thus the aim of the present section is to 
verify the mean-field predictions Eq.~(\ref{eq.finite-correction}) and Eq.~(\ref{eq.int-shift}).

  At any temperature $T$, the $N$ number of bosons are distributed in the energy levels $E_{nl}$ according to the Bose distribution,
\begin{equation}
 f(E_{nl}) = \frac{1}{e^{\beta(E_{nl}-\mu)} -1}
\label{distribution}
\end{equation}
where $\beta=\frac{1}{k_BT}$ and $\mu$ is the chemical potential. The particle number is given by
\begin{equation}
 N = \sum_{n=0}^{\infty} \sum_{l=0}^{\infty} \frac{(2l+1)}{e^{\beta(E_{nl}-\mu)} -1}
 \label{Nsum}
\end{equation}
The average energy $E(N,T)$ of the system at $T$ is given by
\begin{equation}
 E(N,T)=\sum_{n=0}^{\infty} \sum_{l=0}^{\infty}(2l+1)f(E_{nl})E_{nl}
 \label{Esum}
\end{equation}
The heat capacity $C_N(T)$ at a fixed $N$ is calculated by taking the partial derivative of $E(N,T)$ in Eq.~(\ref{Esum}) 
with respect to $T$ as
\begin{equation}
 C_{N}(T) = \frac{\partial E(N,T)}{\partial T}{\Big |}_{N}
\label{spheat}
\end{equation}
As $\mu$ is a function of $T$, we compute the differentiation and obtain
\begin{eqnarray}
C_{N}(T) = \beta\sum_{n=0}^{\infty}\sum_{l=0}^{\infty} \frac{(2l+1)E_{nl}\exp{(\beta(E_{nl}-\mu))}}{(\exp{(\beta(E_{nl}-\mu))}-1)^2}     \nonumber \\
 \times {\Big[}\frac{E_{nl}-\mu}{T}+\frac{\partial \mu}{\partial T}{\Big]},    \nonumber \\
\label{spheatform}
\end{eqnarray}
where 
\begin{equation}
\begin{array}{cl}
\frac{\partial \mu}{\partial T} = & \\
- \frac{\sum_{m=0,l_m=0}^{\infty}(2l_m+1)(E_{ml_m}-\mu)\exp{(\beta(E_{ml_m}-\mu))}(f(E_{ml_m}))^2}{T\sum_{p=0,l_p=0}^{\infty}(2l_p+1)\exp{(\beta(E_{pl_p}-\mu))}(f(E_{pl_p}))^2} &.  
\end{array}
\label{delmudelt}
\end{equation}
\begin{figure}[hbpt]
\vspace{-10pt}
\centerline{
\hspace{-3.3mm}
\rotatebox{0}{\epsfxsize=8.8cm\epsfbox{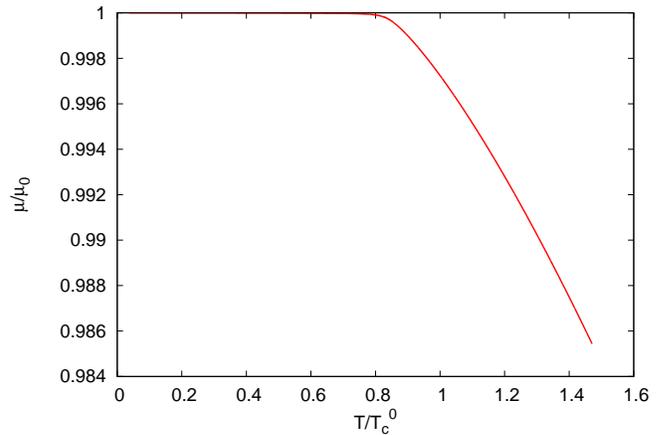}}}
\caption{(color online) Plot of the chemical potential $\mu$ for $N=500$ bosons in the harmonic trap.}
\label{fig.mu_harmonic}
\end{figure}

\begin{figure}[hbpt]
\vspace{-10pt}
\centerline{
\hspace{-3.3mm}
\rotatebox{0}{\epsfxsize=8.8cm\epsfbox{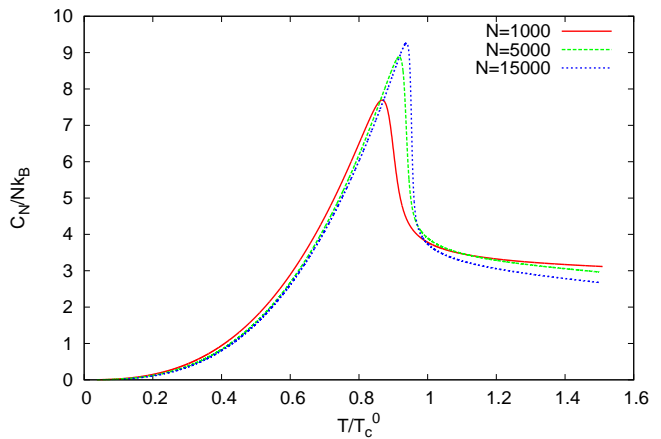}}}
\caption{(color online) Plot of the specific heat $C_N/Nk_B$ for various number of  bosons in the harmonic trap.}
\label{fig.cv_harmonic}
\end{figure}

Next we calculate $C_N(T)$ as a function of $T$ for different values of $N$ by the following procedure. We evaluate the 
quantity \
\begin{equation}
F(\mu, T) = \sum_{n,l=0}^{nmax,lmax} (2l+1)f(E_{nl}) - N
\label{eq.mu}
\end{equation} 
taking a certain number of levels. We check for convergence by increasing the maximum number of levels until $\mu(T)$ is converged. 
In Fig.~\ref{fig.mu_harmonic} we plot $\mu/\mu_0$ ($\mu_0$ is the chemical potential at $T=0$) as a function of $T/T_c^0$ 
($T_c^0$ is the critical temperature in the thermodynamic limit given by $k_BT_c^0=0.94 \hbar \omega N^{1/3}$~\cite{Dalfovo}).
Upto the condensation temperature $\mu$ remains practically constant beyond which it starts to decrease rapidly. 
After getting $\mu(T)$, we calculate $\frac{\partial \mu}{\partial T}$ by Eq.~(\ref{delmudelt}) and next obtain $C_N(T)$ 
from Eq.~(\ref{spheatform}). In Fig.~\ref{fig.cv_harmonic} we show the results of our numerical calculation for different 
values of $N$. For $N=1000$, $C_N(T)$ smoothly changes as a function of $T/T_c^0$. So it is difficult to define 
the critical temperature for condensation. However we observe sharp change near $T=T_c$ for $N=15000$. Thus we define the 
transition temperature $T_c$ at the maximum of the curve $C_N(T)$ for finite $N$ as
\begin{equation}
 \frac{\partial C_N}{\partial T}{\Big |}_{T=T_c}=0.
 \label{def_tc}
\end{equation} 
The calculated values of $T_c$ as a function of $N$ and gas parameter $na_s^3$ are shown in Table~\ref{t-tc-harmonic}.
Our results are systematically lower than the values of Ref.~\cite{Pearson} for hard sphere bosons. The value of $T_c$ smoothly increases with $N$ as expected. It is to be noted that our calculated transition temperature $T_c$ 
slowly approaches to the experimentally measured value $T_c=0.94(5)T_c^0$ (for $N=40000$ $^{87}$Rb atoms)~\cite{Enshar}. 
Thus our numerical results correctly reproduce the condensate temperature 
of the finite sized condensate. 

\begin{center}
\begin{table}[!]
\caption{Transition temperature $T_c$ for different values of gas parameter $na_s^3$.}
\begin{tabular}{|l|l|l|}
 \hline
N &  $na_s^3$ & $\frac{T_c}{T_c^0}$\\ \hline
200 & $1.62 \times 10^{-5}$ & 0.8032\\ \hline
1000 & $8.118 \times 10^{-5}$ & 0.8676\\ \hline
3000 & $2.435 \times 10^{-4}$ & 0.9089\\ \hline
5000 & $4.059 \times 10^{-4}$ & 0.9172\\ \hline
7000 & $5.683 \times 10^{-4}$ & 0.9233\\\hline
9000 & $7.306 \times 10^{-4}$ & 0.9282\\ \hline
11000& $8.93 \times 10^{-4}$ & 0.9312 \\ \hline
13000& $1.055 \times 10^{-3}$ & 0.9349 \\ \hline
15000& $1.218 \times 10^{-3}$ & 0.9369\\ \hline
\end{tabular}
\label{t-tc-harmonic}
\end{table}
\end{center}

\begin{figure}[hbpt]
\vspace{-10pt}
\centerline{
\hspace{-3.3mm}
\rotatebox{0}{\epsfxsize=8.8cm\epsfbox{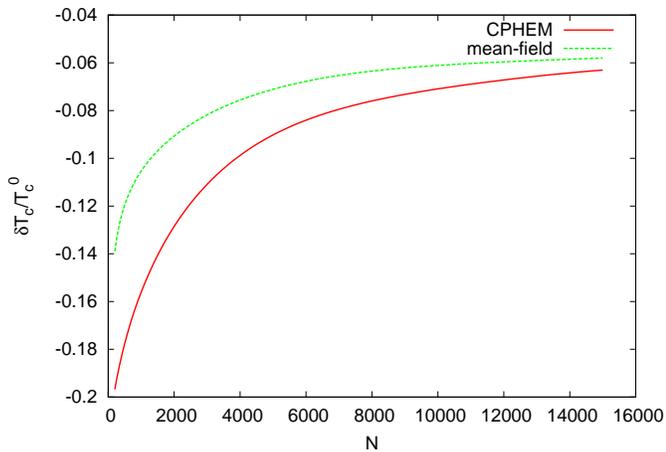}}}
\caption{(color online) Plot of the total relative shift in transition temperature $\frac{\delta T_c}{T_c^0}$ 
of finite number of interacting bosons in the harmonic trap as a function of $N$. Our CPHEM results are presented by the red continuous curve. And the green dashed curve represents the mean-field prediction.}
\label{fig.tot-shift}
\end{figure} 
Earlier in a preliminary study we have shown that for finite sized interacting systems the values of $T_c$ calculated by our CPHEM 
considerably deviates from $T_c^0$~\cite{Satadal}. Now to study it in more systematic way we plot the total relative shift 
$\frac{\delta T_c}{T_c^0} = \frac{T_c-T_c^0}{T_c^0}$ in Fig.~\ref{fig.tot-shift}.   For comparison we also plot the mean-field prediction for the relative 
total shift obtained by adding Eq.(\ref{eq.finite-correction}) and Eq.(\ref{eq.int-shift}) in the same figure. We observe that our CPHEM results are systematically lower than mean-field predictions. However for large $N$ limit our CPHEM results slowly approaches 
the mean-field results. It is justified since in the large $N$ limit the repulsive BEC becomes less correlated and the mean-field 
result becomes a better approximation. 

\begin{figure}[hbpt]
\vspace{-10pt}
\centerline{
\hspace{-3.3mm}
\rotatebox{0}{\epsfxsize=8.8cm\epsfbox{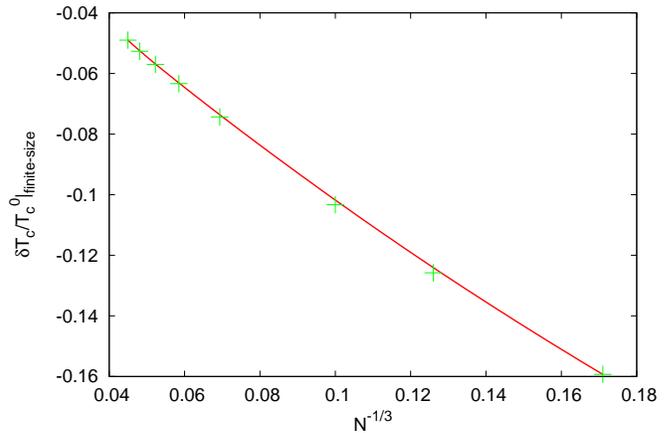}}}
\caption{(color online) Plot of the relative shift in transition temperature $\frac{\delta T_c}{T_c^0}|_{finite-size}$ 
due to finite number of non-interacting bososns in the harmonic trap as a function of $N^{-1/3}$. The green plus signs indicate the position of numerical data points. The red continuous line is drawn by fitting the points.}
\label{fig.finitesize}
\end{figure}

\begin{figure}[hbpt]
\vspace{-10pt}
\centerline{
\hspace{-3.3mm}
\rotatebox{0}{\epsfxsize=8.8cm\epsfbox{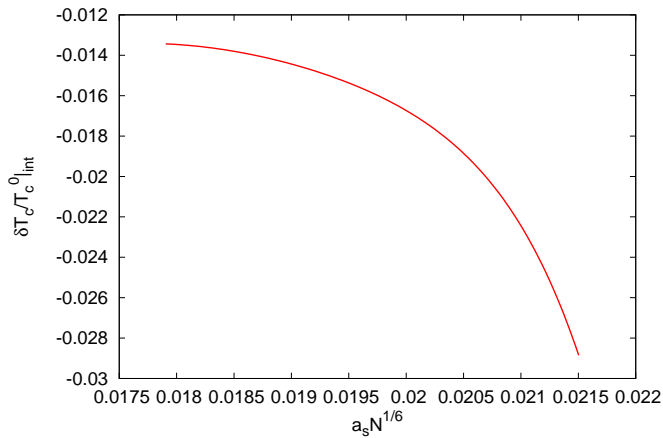}}}
\caption{(color online) Plot of the relative shift in transition temperature $\frac{\delta T_c}{T_c^0}|_{int}$ 
due to the inter-atomic interaction of finite number of bosons in the harmonic trap as a function of $a_sN^{1/6}$. 
}
\label{fig.interaction}
\end{figure}

Next to check the $N^{-1/3}$ dependence of the finite size correction predicted from the mean-field theory [Eq.~(\ref{eq.finite-correction})] we calculate the relative shift in $T_c$ due to finite size effect 
$\frac{\delta T_c}{T_c^0}|_{finite-size} = \frac{T_c^{finite}-T_c^0}{T_c^0}$ and plot it as a function of $N^{-1/3}$ in Fig.~\ref{fig.finitesize}. Here $T_c^{finite}$ is our numerically calculated value of the transition temperature 
for non-interacting bosons in pure harmonic trap. As 
expected, with gradual increase in particle number the correction term will decrease gradually and finally 
in the thermodynamic limit it will vanish.  Then to check the effect of interatomic interaction 
we calculate $\frac{\delta T_c}{T_c^0}|_{int}=\frac{T_c-T_c^{finite}}{T_c^{0}}$  and 
plot in Fig~\ref{fig.interaction} as a function of $a_s N^{1/6}$. Direct comparison with Eq.~(\ref{eq.int-shift}) 
shows significant difference between the mean-field result and our many-body result. It is also to be noted that our 
present results consider the interaction regime which is parametrized by $a_s N^{1/6} \simeq 0.022$. This is to make 
sure that our system of Bose gas is dilute, weakly interacting and 
our choice of two-body correlated basis function remains justified.
However in our correlated many-body method we observe that 
$\frac{\delta T_c}{T_c^0}|_{int}$ gradually goes down with increase in $a_sN^{1/6}$, although the rate is slow. We explain this
as the effect of interatomic correlation which makes the interaction energy more negative.    

\subsection{BEC in anharmonic trap and calculation of several thermodynamic properties.}

\begin{figure}[hbpt]
\vspace{-10pt}
\centerline{
\hspace{-3.3mm}
\rotatebox{0}{\epsfxsize=8.8cm\epsfbox{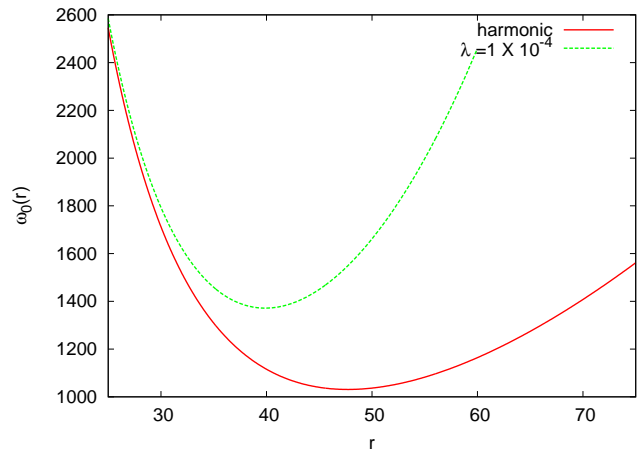}}}
\caption{(color online) Plot of the effective potential $\omega_0(r)$ for $N=500$ bosons in the harmonic trap ($\lambda=0$), anharmonic trap ($\lambda = 1 \times 10^{-4}$).}
\label{fig.pot}
\end{figure}

The potential wells that trap the atoms in the experiments are generally approximated as harmonic oscillator potential. 
However the condensation temperature strongly depends on the trap geometry. In the earlier calculations the 
power-law traps~\cite{Pearson1, Bayindir, Zobay, Jaouadi} are presented as lowering the number of dimensions increases the transition temperature. In the present section we consider $3D$ quadratic plus quartic confining potential as $V(r) = \frac{1}{2} m \omega r^2 +  \lambda r^4$ where $\lambda$ 
is a controllable parameter. For $\lambda > 0$, the quartic confinement becomes more tight and one can switch from the harmonic 
to anharmonic trap by tuning $\lambda$. In the experiments this is achieved by tuning the laser intensity. The choice of anharmonic 
trap also satisfies the purpose of adiabatic formation of BEC. The adiabatic cooling of the gas was proposed as when the trap frequency changes from $\omega$ to $\omega^{\prime}$, keeping the entropy and particle number constant, $T$ chages to 
$\frac{\omega^{\prime}}{\omega}T$~\cite{Houbiers}. 
Thus by varying the shape of the confining trap it is possible to cool the system from high 
temperature phase without condensation down to temperatures below $T_c$ with large condensation fraction in the ground state~\cite{Pinkse, DStamper}. 
Our next findings also indicate that by tuning $\lambda$ slowly compared to the internal equilibration time we can vary 
the shape of the trap adiabatically and thereby bring the system (which is initially above $T_c$) below the transition temperature 
$T_c$. In the experiments the quartic confinement is created by blue detuned gaussian laser directed along the axis and the 
strength of the quartic confinement was $\lambda \sim 10^{-3}$. Thus in our present study we keep $|\lambda|<<1$. In 
Fig.~\ref{fig.pot} we plot the many-body effective potential $\omega_0(r)$ for a fixed particle $N=500$ for $\lambda=0$ 
and $\lambda = 1 \times 10^{-4}$. The trap becomes more tight for $\lambda > 0$, it increases the transition temperature and 
the situation becomes more favorable for BEC. Thus the finite-sized non-ideal Bose gas in anharmonic trap
 deserves special attention as it is not
considered yet. This is specially interesting as it is directly relevant to possible future experimental set up.  

\begin{figure}[hbpt]
\vspace{-10pt}
\centerline{
\hspace{-3.3mm}
\rotatebox{0}{\epsfxsize=8.8cm\epsfbox{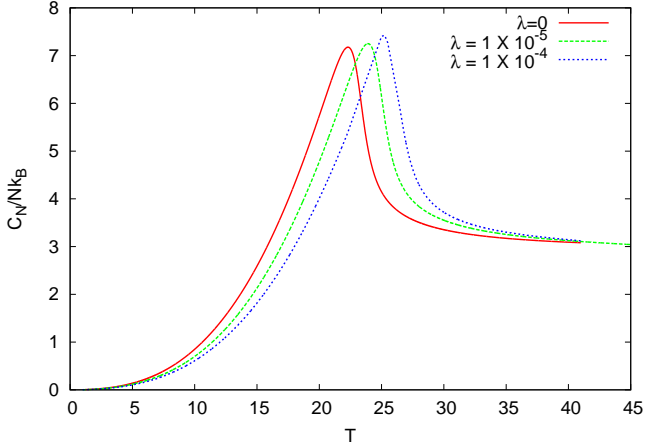}}}
\caption{(color online) Plot of the specific heat $C_N/Nk_B$ for $N=500$ in the anharmonic trap with 
different anharmonicity strength.}
\label{fig.cv_anharmonic}
\end{figure}

\begin{figure}[hbpt]
\vspace{-10pt}
\centerline{
\hspace{-3.3mm}
\rotatebox{0}{\epsfxsize=8.8cm\epsfbox{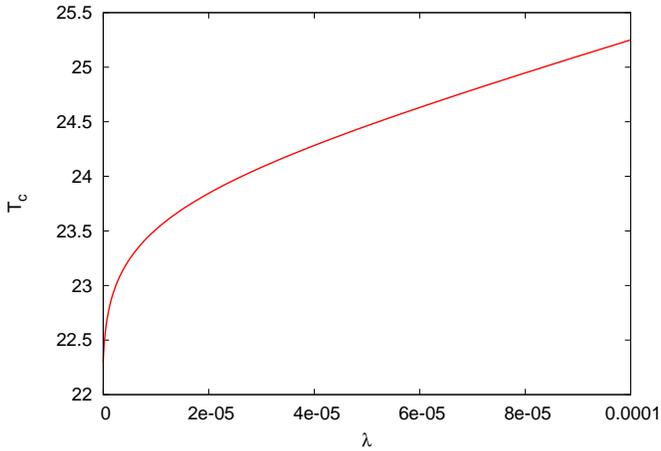}}}
\caption{(color online) Plot of the critical temperature $T_c$ as a 
function of $\lambda$ in the harmonic plus quartic trap for $N = 500$.}
\label{fig.tc_n=500}
\end{figure}

\begin{figure}[hbpt]
\vspace{-10pt}
\centerline{
\hspace{-3.3mm}
\rotatebox{0}{\epsfxsize=8.8cm\epsfbox{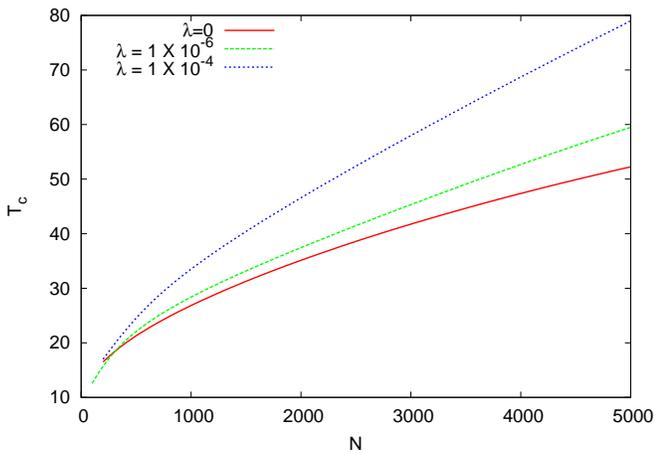}}}
\caption{(color online) Plot of the critical temperature $T_c$ as a function of $N$ in 
the harmonic and anharmonic trap ($\lambda = 1 \times 10^{-6}$ and $\lambda = 1 \times 10^{-4}$).}
\label{fig.tc_anh=1.d-4}
\end{figure}

As before we calculate the energy levels $E_{nl}$ in the effective potential and calculate the specific heat. In Fig.~\ref{fig.cv_anharmonic} we plot $C_N(T)$ as a function of $T$ for different values of $\lambda$. One can observe that 
for pure harmonic trap ($\lambda =0$) and weakly anharmonic (small $\lambda$) trap $C_N(T)$ smoothly  changes ruling out 
any kind of phase transition. However for $\lambda = 1 \times 10^{-4}$, $C_N(T)$ shows sharp transition near $T=T_c$ which 
signals the possibility of pseudo-transition in very tight trap. As before we calculate the transition temperature from 
the position of the peak of $C_N(T)$ and in Fig.~\ref{fig.tc_n=500} we plot $T_c$ as a function of $\lambda$ for $N=500$. 
We observe that $T_c$ increases steadily as the trap becomes more tight. This is in agreement with earlier calculations for 
tight traps which were modelled either by pure quartic potential or by power-law potential. Earlier several predictions about
the amount of increment of $T_c$ have been made for IBG in different trap geometry~\cite{Jaouadi, Gautam}. 
Here, as mentioned above, we are considering interacting bosons in 
harmonic plus quartic trap which is relevant to the adiabatic cooling. We observe that even for a small number of
bosons $N$ in the trap, the transition temperature $T_c$ in the case of anharmonic trap 
with $\lambda=1 \times 10^{-4}$ is almost $10\%$ higher than 
that of the isotropic harmonic trap. Also the amount of increment increases steadily with the number of bosons $N$ in the trap 
as evident from Fig.~\ref{fig.tc_anh=1.d-4}. For $N=5000$ in the anharmonic trap with $\lambda =  1 \times 10^{-4}$, $T_c$ is found
to increase approximately by $37.36\%$ from that of the isotropic harmonic trap. Thus our study not only supports the 
possibility of adiabatic formation of BEC, we for the first time attempt to predict the amount of increment of $T_c$ in an anharmonic trap. Thus our study reveals that it is also possible to adiabatically cool the system by tuning the intensity of 
the laser beam used for creating the trap. 

\begin{figure}[hbpt]
\vspace{-10pt}
\centerline{
\hspace{-3.3mm}
\rotatebox{0}{\epsfxsize=8.8cm\epsfbox{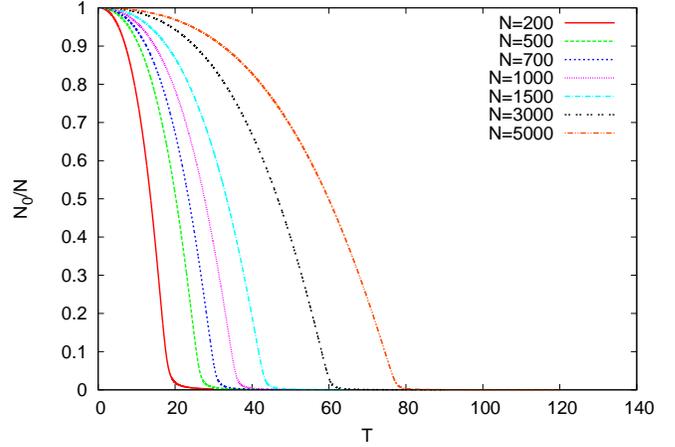}}}
\caption{(color online) Plot of the condensate fraction for various $N$  in the anharmonic trap with 
the anharmonicity strength $\lambda=1 \times 10^{-4}$.}
\label{fig.confrac_anharmonic}
\end{figure}

\begin{figure}[hbpt]
\vspace{-10pt}
\centerline{
\hspace{-3.3mm}
\rotatebox{0}{\epsfxsize=8.8cm\epsfbox{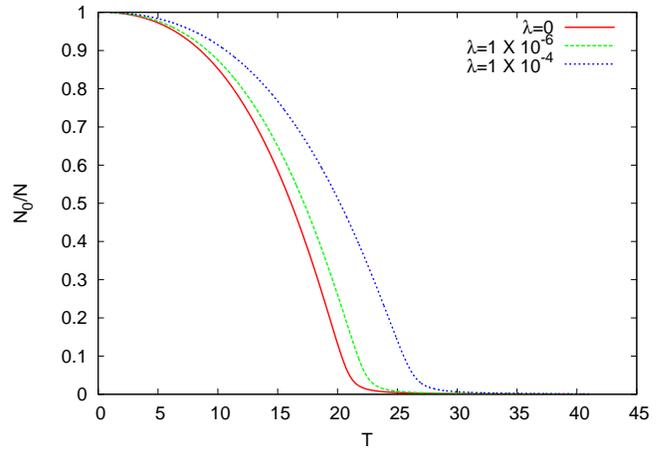}}}
\caption{(color online) Plot of the condensate fraction for $N=500$  in the harmonic and anharmonic trap with 
the various anharmonicity strengths $\lambda$.}
\label{fig.confrac_N=500}
\end{figure}

Next we compute another important thermodynamic quantity - the condensate fraction $N_0/N$. 
In an earlier attempt~\cite{Satadal} we have already
 studied the temperature dependence of the condensate fraction of the interacting Bose gas with a finite number of bosons in the 
 isotropic harmonic trap. We have observed that the main difference of the signature  of BEC in such finite sized system from the 
 infinite limit lies in the shifted and smeared out onset of macroscopic population of the ground state. Here we are
 interested in the temperature variation of the ground state occupation in a harmonic plus quartic potential. By inserting the values of the chemical potential $\mu$ in Eq.~(\ref{eq.mu}) we can calculate the ground state population $N_0$ at a particular 
 temperature and for a fixed $N$. In Fig.~\ref{fig.confrac_anharmonic} we plot $\frac{N_0}{N}$ as a function of $T$ for different
 $N$ in anharmonic trap with $\lambda= 1 \times 10^{-4}$. We observe that as $N$ increases the onset of macroscopic population in 
 the ground state becomes more sudden. Qualitatively this looks very similar to the harmonic trap case. However the onset of a 
 macroscopic ground state occupation occurs at higher temperature for the anharmonic trap. To illustrate the point further, we 
 plot the condensate fraction $\frac{N_0}{N}$ as function of $T$ for $N=500$ bosons in harmonic and anharmonic trap with different 
 $\lambda$ values in Fig.~\ref{fig.confrac_N=500}. It is seen that at any finite $T$ the ground state occupation increases with 
 $\lambda$. This is consistent with the above discussion as well as earlier theoretical calculations in the quartic trap~\cite{Karabulut}.
\section{Conclusion}
In conclusion, we have thoroughly studied several measures of BEC like specific heat, critical temperature and condensate fraction 
in both the isotropic harmonic and the anharmonic trap. We have precisely determined $T_c$ of the interacting bosons in pure harmonic trap and compared it 
with the mean-field results. The detailed study of the finite size correction due to finite number of atoms in the trap and 
also the correction to $T_c$ due to interaction are two major issues addressed in the present manuscript. Next we extend our
correlated many-body method to calculate $T_c$ in the case of quadratic with an additional quartic trap. The choices of 
anharmonic trap also satisfies the purpose of adiabatic formation of BEC. Simply by varying the confining trap it is possible 
to condense the Bose gas at higher $T_c$ which becomes more favourable experimentally. In our theoretical study the anharmonic 
trap is modelled as $V(r) = \frac{1}{2} m \omega^2 r^2 + \lambda r^4$, where $\lambda$ is treated as a controllable parameter.
Experimentally this can be achieved by tuning the intensity of laser beam. Our study reveals significant increase of $T_c$ for a
finite number of interacting bosons with strong anharmonicity.

At present we are not able to compare our theoretical results with the experimental findings. However the tuning of $\lambda$
parameter basically resemble the experimental situation where the shape of the external trap is changed by tuning the laser intensity. Thus our present theoretical results may provide useful information for future experiments.    
\hspace*{.5cm}
\begin{center}
{\large{\bf{Acknowledgements}}}
\end{center}
SKH acknowledges a senior research fellowship from CSIR, India (File No. 08/561(0001)/2010-EMR-I).
SB and TKD acknowledge the financial support of University Grants Commission, India under a Major Research Project [F.N0. 40-439/2011(SR)].

\end{document}